\documentclass[onecolumn, 11pt, draftclsnofoot, letterpaper]{IEEEtran}
\ifCLASSINFOpdf
  \usepackage[dvips]{graphicx}
 \else
 	\usepackage[dvips]{graphicx}
  \fi
\usepackage{amsmath, amsfonts, amssymb, xcolor, lipsum}
\usepackage{mleftright}
\usepackage{cite}

\usepackage{balance}
\DeclareMathOperator*{\argmin}{\arg\!\min}
\DeclareMathOperator*{\argmax}{\arg\!\max}

\begin{document}
\title{Physical-Layer Network Coding with Multiple Antennas: An Enabling Technology for Smart Cities}

\author{\IEEEauthorblockN{Vaibhav Kumar\IEEEauthorrefmark{1},
Barry Cardiff\IEEEauthorrefmark{2}, and Mark F. Flanagan\IEEEauthorrefmark{3}}\\
\IEEEauthorblockA{School of Electrical and Electronic Engineering \\ University College Dublin, Belfield, Dublin 4, Ireland\\
Email: \IEEEauthorrefmark{1}vaibhav.kumar@ucdconnect.ie,
\IEEEauthorrefmark{2}barry.cardiff@ucd.ie,
\IEEEauthorrefmark{3}mark.flanagan@ieee.org}}
\maketitle

\begin{abstract}
Efficient heterogeneous communication technologies are critical components to provide flawless connectivity in smart cities. The proliferation of wireless technologies, services and communication devices has created the need for green and spectrally efficient communication technologies. Physical-layer network coding (PNC) is now well-known as a potential candidate for delay-sensitive and spectrally efficient communication applications, especially in bidirectional relaying, and is therefore well-suited for smart city applications. In this paper, we provide a brief introduction to PNC and the associated distance shortening phenomenon which occurs at the relay. We discuss the issues with existing schemes that mitigate the deleterious effect of distance shortening, and we propose simple and effective solutions based on the use of multiple antenna systems. Simulation results confirm that full diversity order can be achieved in a PNC system by using antenna selection schemes based on the Euclidean distance metric. 
\end{abstract}


% \fi
\IEEEpeerreviewmaketitle

\section{Introduction}
The objective of a \emph{smart city} is to develop the infrastructure in the city in such a manner that it gives a high-quality standard of living to residents, a sustainable environment, and smart solutions to existing problems \cite{SmartCities}. The essential infrastructure components in the smart city would consist of an adequate electricity and water supply, a smart waste management system, an ultra-modern and efficient public transport system, affordable housing, robust and state-of-the-art communication connectivity, safety and security to all citizens especially during natural and man-made disasters, and high-quality health and educational facilities. 

With the rising rural-to-urban demographic shift, more people are migrating to cities; $70\%$ of the world's total population will be living in cities by 2050, according to an estimate by the United Nations. This surge in the number of inhabitants will give rise to new challenges for communication infrastructure design in these cities, as these communication networks will enable  interoperable access and interconnection among heterogeneous smart city objects~\cite{ComMag17}. Examples of the involved communication technologies for efficient service delivery in these high-tech cities include wireless sensor networks (WSN), device-to-device (D2D) communication, vehicle-to-vehicle (V2V) communication, software defined networks (SDN), mm-wave communications, internet-of-things (IoT) \cite{IoT}, spectrum sharing systems, energy harvesting systems \cite{Talla} and backscatter communications \cite{Backscatter}. 

It has been forecasted that the total number of connected devices by the end of 2020 will reach the 50 billion mark, and mobile data traffic will reach 49 EB per month (1 EB = $10^{18}$ bytes) by 2021 \cite{Cisco}. This growth in data traffic associated with a plethora of data-intensive applications puts an enormous demand on the available licensed as well as unlicensed frequency bands. The limited available bandwidth, together with the inefficiencies of current spectrum allocation strategies, necessitates new communication paradigms to exploit the available spectrum. Wireless PNC \cite{HotTopic} is a potential candidate to enable efficient spectrum usage, possessing desirable properties such as delay reduction and throughput enhancement in a bidirectional relaying scenario. These inherent properties make PNC an excellent candidate for many of the applications in a smart city such as in a WSN where access points exchange information via a mobile relay, D2D communications where a relay helps two smart devices to communicate with each other, and two earth-stations in different cities communicating via a satellite. 
%In the present paper we provide a brief introduction to PNC, some of the associated problems and feasible solutions. In particular, we emphasize on the distance shortening phenomenon and its solution using multiple antenna PNC system.

Different aspects of PNC relating to communication theory, information theory, wireless networking, and finite and infinite field PNC, as well as synchronization issues and the use of PNC for passive optical networks were discussed in \cite{PNC_Tutorial, Primer, FieldPNC}. The first software radio based implementation of PNC was reported in \cite{Katti}, together with a discussion on related problems and solutions. A comprehensive discussion and performance comparison in terms of rates in bits per channel use (bpcu) of lattice-code-based PNC, analog PNC, wireless broadcast network coding and routing-based bidirectional relaying has been presented in \cite{Gastpar}; it was shown therein that PNC systems based on lattice coding can achieve the upper bound on rate per channel use for higher transmit power levels in an interference-limited scenario. Building on~\cite{Gastpar}, an algebraic approach to PNC was presented in \cite{AlgebraicPNC} via the use of a nested lattice. PNC-based random access has been discussed in \cite{Goseling} for a two-user network: here, in each round, transmitters choose a random state of being either active or idle, and the corresponding receivers decode a linear combination of messages in each round; a sequence of linear combinations is thus obtained, leading to the eventual recovery of all of the original data packets. A Gaussian integer based formulation for complex linear PNC was discussed in \cite{ComplexPNC} for a bidirectional relaying scenario, where the symbols transmitted from users and the corresponding network-coded symbol at the relay are the elements of a finite field of Gaussian integers.

A general framework for the symbol-error-rate (SER) performance analysis of PNC systems in AWGN channel has been presented in \cite{SER}. The phenomenon of distance shortening, which will be explained in detail later in this paper, is one of the major bottlenecks in wireless PNC in the context of bidirectional relaying. To solve this problem, a number of \emph{adaptive physical-layer network coding} (ANC) schemes have been proposed \cite{Akino, RajanPNC, AdaptiveOFDM}, where the relay adaptively selects the network map that offers the best performance based on the channel conditions. A detailed introduction to wireless multi-way relaying using ANC was presented in \cite{MultiWay}. It has been shown in \cite{LatinSquares} that every valid network map can be represented by a Latin square and this relationship can be used to obtain network maps with optimized intercluster distance profiles. Although ANC alleviates the problem of distance shortening in an efficient way, the related system complexity increases significantly due to the required clustering algorithm and the use of non-standard modulation. In \cite{Vaibhav}, a simple antenna selection based solution to mitigate the distance shortening problem was proposed and analyzed, applicable to the case where the users are equipped with multiple antennas and the relay has a single antenna. 

In this paper, we present a brief introduction to wireless PNC, and we explain the distance shortening phenomenon. We propose an antenna selection scheme capable of mitigating this problem, which generalizes the antenna selection aided PNC scheme of \cite{Vaibhav} to the case of multiple relay antennas. We also propose a new joint user \emph{and} relay antenna selection scheme. Finally, we provide an error rate performance comparison between the proposed schemes and an existing strongest-channel-based antenna selection scheme for PNC.

\section{PNC in Two-Way Relay Channel}
\begin{figure}[hbht]
\centering 
\includegraphics[scale=1.1]{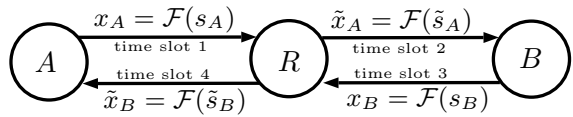}
\caption{Information exchange between users $A$ and $B$ through the conventional approach.}
\label{conventional}
\end{figure}

The broadcast nature of the wireless channel makes it difficult for multiple transmitters to access the radio channel simultaneously, since it can be a difficult task for the intended receiver to distinguish between the useful information and the interference. This destructive \emph{multiple-access interference} (MAI) is one of the major bottlenecks when designing medium access control (MAC) protocols. PNC is a way to exploit this MAI in a constructive manner. To understand the advantage of PNC, consider a scenario where bidirectional information exchange takes place between two users $A$ and $B$. Suppose that there is no line-of-sight (LOS) path available between the users (this assumption can hold due to the nature of the terrain around the users, or due to the presence of severe fading, shadowing and path-loss in the LOS path). The communication between the users take place with the assistance of a relay $R$. We further assume that all three nodes are operating in half-duplex mode. This type of communication channel is popularly known as the two-way relay channel (TWRC). Below we consider two different approaches for the data transfer between the users:
\paragraph{Conventional approach} This approach avoids any kind of MAI, and thus a total of four time slots will be required to exchange two packets (one from each user). In the first time slot, user $A$ will transmit its packet to $R$, and in the second slot $R$ will forward (after processing) the packet received from user $A$ to user $B$. In the third time slot, user $B$ will transmit its packet to the relay $R$, and in the fourth time slot $R$ will forward the packet received from user $B$ to user $A$. The end-to-end performance will depend on the signal processing performed at the relay. In Fig. \ref{conventional}, $s_{m} \in \mathbb{Z}_{M}$ denotes the user message with $m \in \{A, B\}$, $M$ is the modulation order, $\mathbb{Z}_{M} = \{0, 1, \ldots, M-1\}$ denotes the set of integer residues modulo $M$,  $\mathcal{F}(\cdot)$ is the constellation mapper, $x_{m} \in \mathcal{X}$ is the complex modulated symbol ($\mathcal{X}$ is the constellation), $\tilde{s}_{m} \in \mathbb{Z}_{M}$ is the message at the relay after processing and $\tilde{x}_{m} = \mathcal{F}(\tilde{s}_{m}) \in \mathcal{X}$  is the complex symbol transmitted from the relay after processing. 
%\paragraph{Network coded communication} The information exchange between the users can be completed in three time slots if network coding (NC) \cite{HotTopic} is used at the relay. In the first and second time slot user $A$ and $B$ will send their data to the relay respectively.  After receiving the signals from both the users, the relay generates an estimate of the received symbols and then forms a \emph{network coded symbol} (NCS). An example of this NCS can be a simple XOR operation between the symbol estimates i.e., $\tilde{s}_{A} \oplus \tilde{s_{B}} = s_{R}$. The relay then modulates $s_{R}$ to a complex symbol $x_{R} = \mathcal{F}(s_{R})$ and broadcasts it to both the users in the third time slot. Upon receiving the NCS and after demodulation, the users can subtract the symbol they sent to the relay in order to extract the intended information. By completing the information exchange in three time slots rather than four, network coded communication provided a $33 \%$ improvement in throughput. 
%\begin{figure}[hbht]
%\centering 
%\includegraphics[scale=1]{figures/NC.eps}
%\caption{Information exchange between users $A$ and $B$ through network coding approach.}
%\label{NC}
%\end{figure}
 
\begin{figure}[hbht]
\centering 
\includegraphics[scale=1]{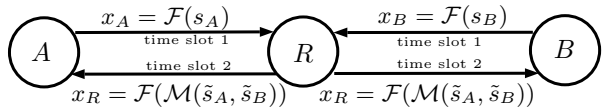}
\caption{Information exchange between users $A$ and $B$ through the PNC approach.}
\label{PNC}
\end{figure}

\paragraph{PNC assisted communication} The wireless communication system assisted with PNC can help to reduce the number of time slots required for the information exchange to two and hence can improve the system throughput by $100\%$. In the first time slot, also termed the multiple-access (MA) phase, both users transmit their information to the relay simultaneously. The relay receives a superposition of the users' signals and performs maximum-likelihood (ML) detection in order to obtain a joint estimate $(\tilde{s}_{A}, \tilde{s}_{B}) \in \mathbb{Z}_{M}^{2}$.  After obtaining the joint estimate, the relay generates a \emph{network coded symbol} (NCS). An example of this NCS can be a simple bitwise exclusive-or (XOR) operation between the symbol estimates, i.e., $\tilde{s}_{A} \oplus \tilde{s}_{B} = s_{R} \in \mathbb{Z}_{M}$ where $\oplus$ denotes bitwise XOR in $\mathbb{Z}_{M}$. The relay then modulates $s_{R}$ to a complex symbol $x_{R} = \mathcal{F}(s_{R})$ and broadcasts it to both users in the next time slot, called the \emph{broadcast} (BC) phase. Using its own message transmitted in the previous MA phase, $A$ can decode the message transmitted from $B$ and vice versa. Fig. \ref{PNC} shows the signals in both MA and BC phases, where $\mathcal{M}(\cdot, \cdot)$ denotes the \emph{network map} used at the relay (also called the \emph{denoising map}).
For successful decoding, this map must follow the \emph{exclusive law} \cite{Akino}:
\begin{equation}
\begin{aligned}
	& \mathcal{M}(s_{1}, s_{2}) \neq \mathcal{M}(s_{1}', s_{2}); \forall s_{1} \neq s_{1}' \in \mathbb{Z}_{M}, s_{2} \in \mathbb{Z}_{M} \\
	& \mathcal{M}(s_{1}, s_{2}) \neq \mathcal{M}(s_{1}, s_{2}'); \forall s_{1} \in \mathbb{Z}_{M}, s_{2} \neq s_{2}' \in \mathbb{Z}_{M} 
\end{aligned} \label{exclusive_law}
\end{equation}
\section{Signal Transmission and Reception in PNC}
In this section, we will elaborate on the signal transmission and reception schemes for the PNC-assisted communication approach.
\subsection{MA phase}
The signal received at the relay $R$ in the MA phase is given by
\begin{equation}
	y_{R} = \sqrt{E_{s}}h_{A} x_{A} + \sqrt{E_{s}}h_{B} x_{B} + n_{R} \label{relay_receive}
\end{equation}
where $E_{s}$ is the energy of the transmitted signal, $h_{m} \sim \mathcal{CN}(0, 1)$ is the complex channel coefficient between user $m \in \{A, B\}$ and relay $R$, and $n_{R} \sim \mathcal{CN}(0, N_{0})$ is the circularly symmetric complex additive white Gaussian noise at the relay. Furthermore, we assume a slow block fading channel between the users and the relay and we also assume that perfect channel state information (CSI) of the $m \rightarrow R$ link is available at the relay only. The ML estimate of the transmitted symbol pair $(x_{A}, x_{B}) \in \mathcal{X}^{2}$ is given by 
\begin{equation}
	(\tilde{x}_{A}, \tilde{x}_{B}) = \argmin_{(x_{A}, x_{B}) \in \mathcal{X}^{2}} \left\vert y_{R} - \sqrt{E_{s}} (h_{A} x_{A} + h_{B} x_{B} ) \right\vert.
\end{equation}
Having this joint estimate $(\tilde{x}_{A}, \tilde{x}_{B}) \in \mathcal{X}^{2}$, the relay calculates $(\tilde{s}_{A}, \tilde{s}_{B}) = \left( \mathcal{F}^{-1}(\tilde{x}_{A}), \mathcal{F}^{-1} (\tilde{x}_{B})\right)$ and then maps this to the network coded symbol $x_{R} \in \mathcal{X}$ using the mapping $\mathcal{M}:\mathbb{Z}_{M}^{2} \rightarrow \mathbb{Z}_{M}$. Table \ref{mapping_table}  shows the mapping for QPSK modulation, where $s_{R} = \tilde{s}_{A} \oplus \tilde{s}_{B} \in \mathbb{Z}_{4}$ and $\oplus$ denotes the bitwise XOR operation in $\mathbb{Z}_{4}$.
\renewcommand{\tabcolsep}{4pt}
\begin{table}[hbht]
\centering
\caption{Example PNC mapping at the relay for QPSK constellation.}
\label{mapping_table}
\begin{tabular}{|c|c|c|c|}
\hline
$(\tilde{s}_{A}, \tilde{s}_{B})$                                                        & $(\tilde{x}_{A}, \tilde{x}_{B})$                                                                                    & $s_{R}$ & $x_{R}$ \\ \hline
\begin{tabular}[c]{@{}c@{}}(0, 0), (1, 1),\\ (2, 2), (3, 3)\end{tabular} & \begin{tabular}[c]{@{}c@{}}$\left(\frac{1+i}{\sqrt{2}}, \frac{1+i}{\sqrt{2}} \right)$, $\left(\frac{-1+i}{\sqrt{2}}, \frac{-1+i}{\sqrt{2}}\right)$,\\ $\left(\frac{-1-i}{\sqrt{2}}, \frac{-1-i}{\sqrt{2}}\right)$, $\left(\frac{1-i}{\sqrt{2}}, \frac{1-i}{\sqrt{2}}\right)$\end{tabular} & 0       & $\frac{1+i}{\sqrt{2}}$     \\ \hline
\begin{tabular}[c]{@{}c@{}}(0, 1), (1, 0),\\ (2, 3), (3, 2)\end{tabular} & \begin{tabular}[c]{@{}c@{}}$\left(\frac{1+i}{\sqrt{2}}, \frac{-1+i}{\sqrt{2}}\right)$, $\left(\frac{-1+i}{\sqrt{2}}, \frac{1+i}{\sqrt{2}}\right)$,\\ $\left(\frac{-1-i}{\sqrt{2}}, \frac{1-i}{\sqrt{2}}\right)$, $\left(\frac{1-i}{\sqrt{2}}, \frac{-1-i}{\sqrt{2}}\right)$\end{tabular} & 1       & $\frac{-1+i}{\sqrt{2}}$    \\ \hline
\begin{tabular}[c]{@{}c@{}}(0, 2), (2,0),\\ (1, 3), (3,1)\end{tabular}   & \begin{tabular}[c]{@{}c@{}}$\left(\frac{1+i}{\sqrt{2}}, \frac{-1-i}{\sqrt{2}}\right)$, $\left(\frac{-1-i}{\sqrt{2}}, \frac{1+i}{\sqrt{2}}\right)$,\\ $\left(\frac{-1+i}{\sqrt{2}}, \frac{1-i}{\sqrt{2}}\right)$, $\left(\frac{1-i}{\sqrt{2}}, \frac{-1+i}{\sqrt{2}}\right)$\end{tabular} & 2       & $\frac{-1-i}{\sqrt{2}}$    \\ \hline
\begin{tabular}[c]{@{}c@{}}(0, 3), (3,0),\\ (1, 2), (2, 1)\end{tabular}  & \begin{tabular}[c]{@{}c@{}}$\left(\frac{1+i}{\sqrt{2}}, \frac{1-i}{\sqrt{2}}\right)$, $\left(\frac{1-i}{\sqrt{2}}, \frac{1+i}{\sqrt{2}}\right)$,\\ $\left(\frac{-1+i}{\sqrt{2}}, \frac{-1-i}{\sqrt{2}}\right)$, $\left(\frac{-1-i}{\sqrt{2}}, \frac{-1+i}{\sqrt{2}}\right)$\end{tabular} & 3       & $\frac{1-i}{\sqrt{2}}$     \\ \hline
\end{tabular}
\end{table}
\subsection{BC phase}
After mapping the joint estimate of the received symbol to the network coded symbol, the relay modulates the network coded symbol and broadcasts to both users. The signal received at nodes $A$ and $B$ can be respectively written as
\begin{equation}
\begin{aligned}
	& y_{A} = \sqrt{E_{s}} h_{A}' x_{R} + n_{A} \\
	& y_{B} = \sqrt{E_{s}} h_{B}' x_{R} + n_{B}
\end{aligned} \label{BC_Phase}
\end{equation}
where $h_{m}' \sim \mathcal{CN}(0, 1)$ is the complex channel fading coefficient between the relay and user $m \in \{ A, B\}$ and $n_{m} \sim \mathcal{CN}(0, N_{0})$ is the noise at the user node. Assuming user $m$ has perfect knowledge of the CSI of the $R \rightarrow m$ link, $A$ can estimate the desired information $s_{B}$ with the help of \eqref{exclusive_law} and the information it has transmitted in the MA phase via 
\begin{equation}
	\hat{s}_{B} = \argmin_{s \in \mathbb{Z}_{M}} \left\vert y_{A} - \sqrt{E_{s}} h_{A}' \mathcal{F}(\mathcal{M}(s_{A}, s))\right\vert.
\end{equation}
User $B$ can decode the message sent by $A$ in a similar fashion. It is interesting to note that the end-to-end error-rate performance of a PNC-assisted communication system is dominated by the error-rate performance in the MA phase, since in the BC phase the communication is similar to a traditional point-to-point communication system; therefore, in the rest of the paper we will focus on the error rate performance in the MA phase.
\section{Clustering and Distance Shortening}
User constellation symbol pairs which are mapped to the same complex number in the network-coded constellation at the relay are said to form a {\emph{cluster}. The error performance in the MA phase will depend on the minimum distance between signal points in different clusters, defined as 
\begin{align}
	& d_{\mathrm{min}}(h_{A}, h_{B})= \min_{\substack{ (s_{A}, s_{B}) \neq (s_{A}', s_{B}') \in \mathbb{Z}_{M}^{2} \\ \mathcal{M}(s_{A}, s_{B}) \neq \mathcal{M}(s_{A}', s_{B}')}} \Big\{ \sqrt{E_{s}} \ \big| h_{A}[\mathcal{F}(s_{A}) - \mathcal{F}(s_{A}')] + h_{B}[\mathcal{F}(s_{B}) - \mathcal{F}(s_{B}')] \big| \Big\} . \label{d_min}
\end{align}
It is clear from \eqref{d_min} that the value of $d_{\mathrm{min}}$ depends on the channels between the users and the relay. It can also be deduced from \eqref{d_min} that there exist values of the ratio $h_{A}/h_{B}$ for which the minimum distance between the clusters becomes zero; such channel states are known as \emph{singular fade states} \cite{LatinSquares}. In general, when the values of the channel coefficients are close to those of the singular fade states, the distance between the clusters is significantly reduced; this phenomenon is called \emph{distance shortening}. 

\begin{figure}[hbht]
\centering 
\includegraphics[scale=0.5]{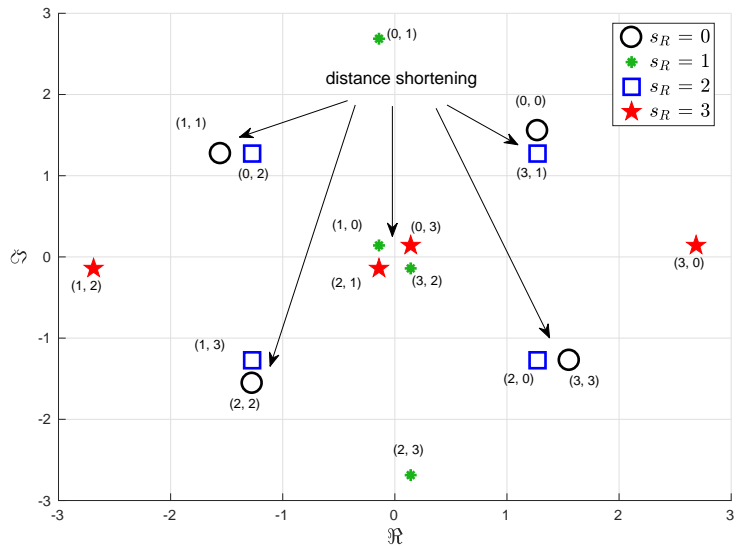}
\caption{Network coded symbols at the relay for QPSK constellation.}
\label{distance_shortening}
\end{figure}

To understand the distance shortening phenomenon, we consider a simple example of one transmission slot where the users transmit their messages using QPSK modulation and the channel coefficients are $h_{A} = (1 + i)/\sqrt{2}$ and $h_{B} = (1 - 0.8i)/\sqrt{2}$. With this combination, the minimum distance between the clusters at the relay becomes very small, which can lead to an incorrect ML estimate at the relay. In Fig. \ref{distance_shortening}, we show the noise-free received signal at the relay, i.e., $h_{A}x_{A} + h_{B}x_{B}$ (assuming $E_{s} = 1$), together with the network-coded symbols, where each 2-tuple in the figure represents $(s_{A}, s_{B})$.

In order to overcome this distance shortening phenomenon, a \emph{closest-neighbor clustering} (CNC) algorithm was proposed in \cite{Akino}, which works by successively placing the closest pair of network-coded symbols in the same cluster provided that the pair satisfies the exclusive law. The CNC algorithm results in a network map having the largest intercluster distance profile for the considered value of $h_{A} / h_{B}$. It has been shown in \cite{Akino} that for a 4-ary modulation scheme in MA phase, CNC may result in a 5-ary network map, and therefore a non-standard 5-ary modulation scheme will be required for the BC phase under certain channel conditions. Also, the use of such an adaptive network coding scheme increases the system complexity significantly, due to the adaptive selection of the relay mapping based on channel conditions; this method can also incur a sacrifice in the reliability at the BC stage due to the increased cardinality of the relay's transmit constellation.

%In this paper, we show that the use of multiple antennas in PNC system can mitigate the deleterious effect of the distance shortening phenomenon and also overcome the drawbacks of ANC. 
%To mitigate the deleterious effect of the distance shortening phenomenon and to overcome the drawbacks of ANC, we propose new antenna selection schemes for multiple-antenna based PNC system.
\section{Mitigating Distance Shortening in PNC using Multiple-Antenna Techniques} 
The system model for PNC with multiple user antennas and multiple relay antennas is shown in Fig \ref{SystemModel}; users $A$ and $B$ are equipped with $N_{A} \ (>1)$ and $N_{B} \ (> 1)$ antennas respectively, while the relay is equipped with $N_{R} \geq 1$ antennas. 
\begin{figure}[hbht]
\centering
\includegraphics[scale=1.4]{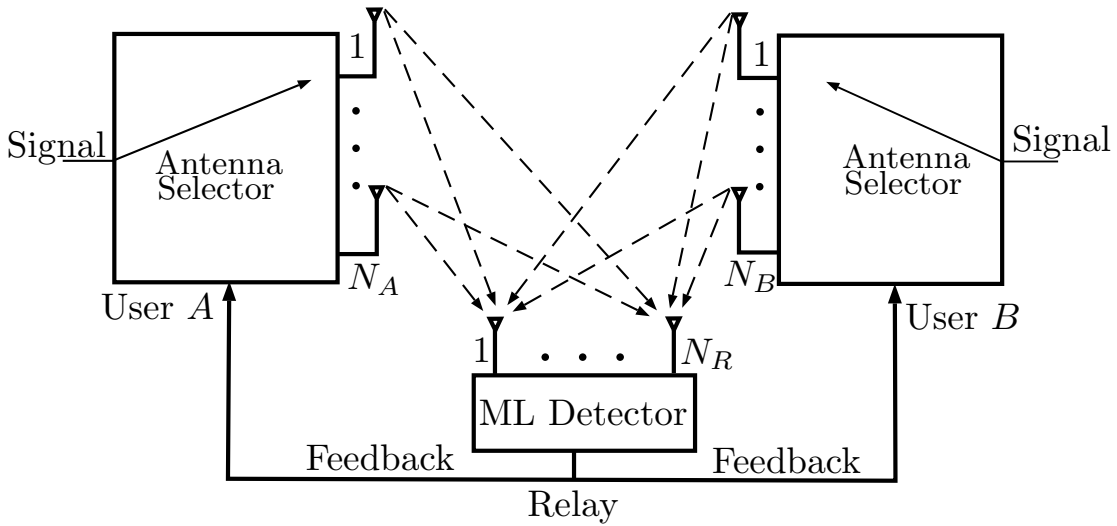}
\caption{System model for PNC with multiple user and multiple relay antennas.}
\label{SystemModel}
\end{figure}
During every transmission, only one of the antennas from each user is used for transmission, and the choice of antennas is based on feedback received from the relay. The channel $h_{m, i, j}, m \in \{A, B\}$ between the $i^{\mathrm{th}}$ antenna of user $m$ and the $j^{th}$ antenna of relay $R$ is modeled as slow Rayleigh fading where perfect CSI is available at $R$ only, with $1 \leq i \leq N_{m}$ and $1 \leq j \leq N_{R}$. This system model has been analyzed in \cite{Huang} assuming BPSK modulation and the implementation of a strongest-channel-based transmit antenna selection (TAS) scheme at each user. In the following subsections, we will discuss this antenna selection scheme, and propose new methods for antenna selection based on the Euclidean distance (ED) metric. 
\subsection{TAS based on strongest channel (TAS1)}
In this scheme \cite{Huang}, the index of the selected antenna at each user is the one having the highest received signal power at the relay; this may be defined as
\begin{equation}
i_{m}^* = \argmax_{1 \leq i \leq N_{m}} \sum_{j = 1}^{N_{R}} |h_{m, i, j}|^{2}. \label{TAS1_index}
\end{equation}
It was shown in \cite{Huang} that a PNC system with the aforementioned TAS scheme achieves a diversity order of $\min(N_{A}, N_{B}) \times N_{R}$ in the MA phase, and closed-form expressions for tight upper and lower bounds on the average bit error rate (BER) were provided. However, in \cite{Vaibhav} it was shown analytically and by simulation that for the case when $N_{R} = 1$, TAS1 fails to exploit the advantage of multiple antennas at the user end to leverage diversity gain for non-binary modulations $(M > 2)$.
\subsection{TAS based on Euclidean distance (TAS2)}
In order to overcome the shortcoming of TAS1, an ED based TAS scheme was proposed in \cite{Vaibhav} for the case when $N_{R} = 1$. In this paper, we describe a more general version of this scheme, applicable to the case when $N_{R} \geq 1$. In TAS2, the transmit antenna of each user is selected such that the minimum ED between the clusters at the relay is maximized. Let $\mathcal{I} = \{(i, j): 1 \leq i \leq N_{A}, 1 \leq j \leq N_{B}\}$ be the set which enumerates all of the possible $n = N_{A} \times N_{B}$ combinations of selecting one antenna from each user. Among these $n$ combinations, the set of transmit antennas that maximizes the minimum ED between the clusters at the relay is obtained as
\begin{align}
	& I_{ED}  = \argmax_{I \in \mathcal{I}} \left\{ \min_{\substack{ \boldsymbol{s, s'} \in \mathbb{Z}_{M}^{2} \\ \mathcal{M}(\boldsymbol{s}) \neq \mathcal{M} (\boldsymbol{s}')}} \left\Vert \boldsymbol{H}_{I} \left(\begin{bmatrix}
	\mathcal{F}(s_{A}) \\ \mathcal{F}(s_{B})
\end{bmatrix} - \begin{bmatrix}
	\mathcal{F}(s_{A}') \\ \mathcal{F}(s_{B}')
\end{bmatrix}\right) \right\Vert^{2}\right\} \label{TAS2_Index}
\end{align}
where $\boldsymbol{H}_{I} = [\boldsymbol{h}_{A, i} \ \boldsymbol{h}_{B, j}] \in \mathbb{C}^{N_{R} \times 2}$, $\boldsymbol{h}_{A, i} = [h_{A, i, 1} \ h_{A, i, 2} \ \ldots \ h_{A, i, N_{R}}]^{T} \in \mathbb{C}^{N_{R} \times 1}$, $\boldsymbol{h}_{B, j} = [h_{B, j, 1} \  h_{B, j, 2} \ \ldots$ $ \ h_{B, i, N_{R}}]^{T} \in \mathbb{C}^{N_{R} \times 1}$, $\boldsymbol{s} = (s_{A}, s_{B})$, $\boldsymbol{s}' = (s_{A}', s_{B}')$ and $\boldsymbol{H}_{I_{ED}} = [\boldsymbol{h}_{A, i^*} \ \boldsymbol{h}_{B, j^*}] \in \mathbb{C}^{N_{R} \times 2}$ is the optimal channel matrix.
\subsection{Joint User and Relay Antenna Selection (JAS) based on ED}
The number of RF chains required for a PNC system with TAS2 is $N_{R} + 2$. In order to reduce the complexity of the PNC system, we here propose a joint user and relay antenna selection scheme which achieves the same diversity order as of TAS2 with only 3 RF chains. In this scheme, antenna selection is performed at the users as well as at the relay, and in such a manner that the minimum ED between the clusters at the relay is maximized. Let $\mathcal{J} = \{(i, j, k): 1 \leq i \leq N_{A}, 1 \leq j \leq N_{B}, 1 \leq k \leq N_{R}\}$ be the set which enumerates all of the possible $p = N_{A} \times N_{R} \times N_{B}$ combinations of selecting one antenna at from each user and the relay. Among these $p$ combinations, the set of antennas that maximizes the minimum ED between the clusters is obtained as
\begin{align}
	& J_{ED}  = \argmax_{J \in \mathcal{J}} \left\{ \min_{\substack{ \boldsymbol{s, s'} \in \mathbb{Z}_{M}^{2} \\ \mathcal{M}(\boldsymbol{s}) \neq \mathcal{M} (\boldsymbol{s}')}} \left\Vert \boldsymbol{H}_{J} \left(\begin{bmatrix}
	\mathcal{F}(s_{A}) \\ \mathcal{F}(s_{B})
\end{bmatrix} - \begin{bmatrix}
	\mathcal{F}(s_{A}') \\ \mathcal{F}(s_{B}')
\end{bmatrix}\right) \right\Vert^{2}\right\} \label{JAS_Index}
\end{align}
where $\boldsymbol{H}_{J}~=~[h_{A, i, k} \ h_{B, j, k}] \in \mathbb{C}^{1 \times 2}$ and $\boldsymbol{H}_{J_{ED}}~=~[h_{A, i^*, k^*} \ h_{B, j^*, k^*}] \in \mathbb{C}^{1 \times 2}$ is the optimal channel vector.
\section{Results and Discussion}
In this section, we present the performance comparison in the MA phase among the three antenna selection schemes discussed in the previous section. We consider the case where both users transmit to the relay using a unit-energy QPSK constellation. We first consider the case when $N_{R} = 1$. In this case the performance of both TAS2 and JAS will be the same and hence we present the symbol error rate (SER) performance comparison between TAS1 and TAS2. It is clear from Fig. \ref{MISO} that the PNC system with TAS1 achieves a diversity order equal to 1 irrespective of the number of antennas at the users (as was proved analytically in \cite{Vaibhav}), as for large $E_{s} / N_{0}$ the SER curve becomes parallel to $(E_{s} / N_{0})^{-1}$. On the other hand, the PNC system with TAS2 outperforms the one with TAS1 while achieving a higher diversity order, and for large $E_{s} / N_{0}$ the SER curve becomes parallel to $(E_{s} / N_{0})^{-\min(N_{A}, N_{B})}$.

\begin{figure}[hbht]
\centering 
\includegraphics[scale=1]{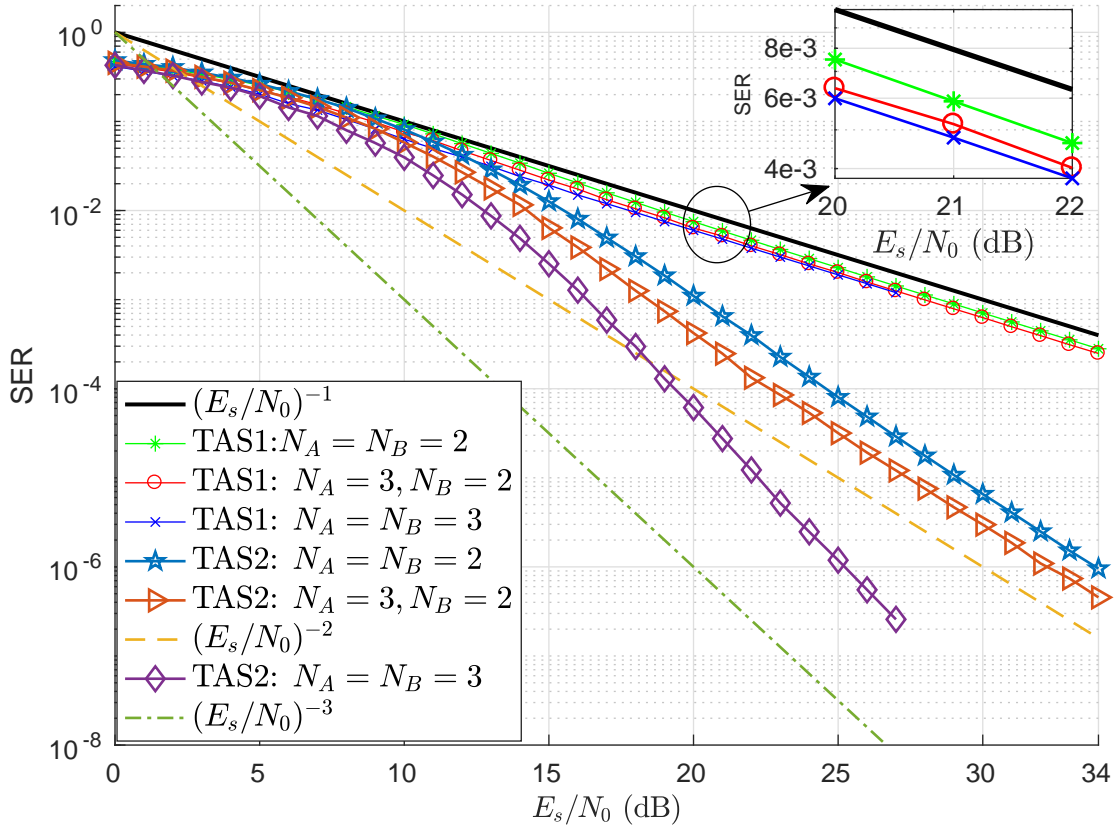}
\caption{Performance comparison of SER for TAS1 and TAS2 in the MA phase for the PNC system.}
\label{MISO}
\end{figure}

Next, we present the performance comparison among TAS1, TAS2 and JAS for the case of $N_{R} > 1$ and QPSK transmission at each user. It is clear from Fig. \ref{MIMO} that for this case, the PNC system with TAS1 achieves a diversity order equal to $N_{R}$ as the SER curve becomes parallel to $(E_{s}/N_{0})^{-N_{R}}$ for large values of $E_{s}/N_{0}$, and the number of transmit antennas at the user has no effect on the diversity order. 
\begin{figure}[hbht]
\centering 
\includegraphics[scale=1]{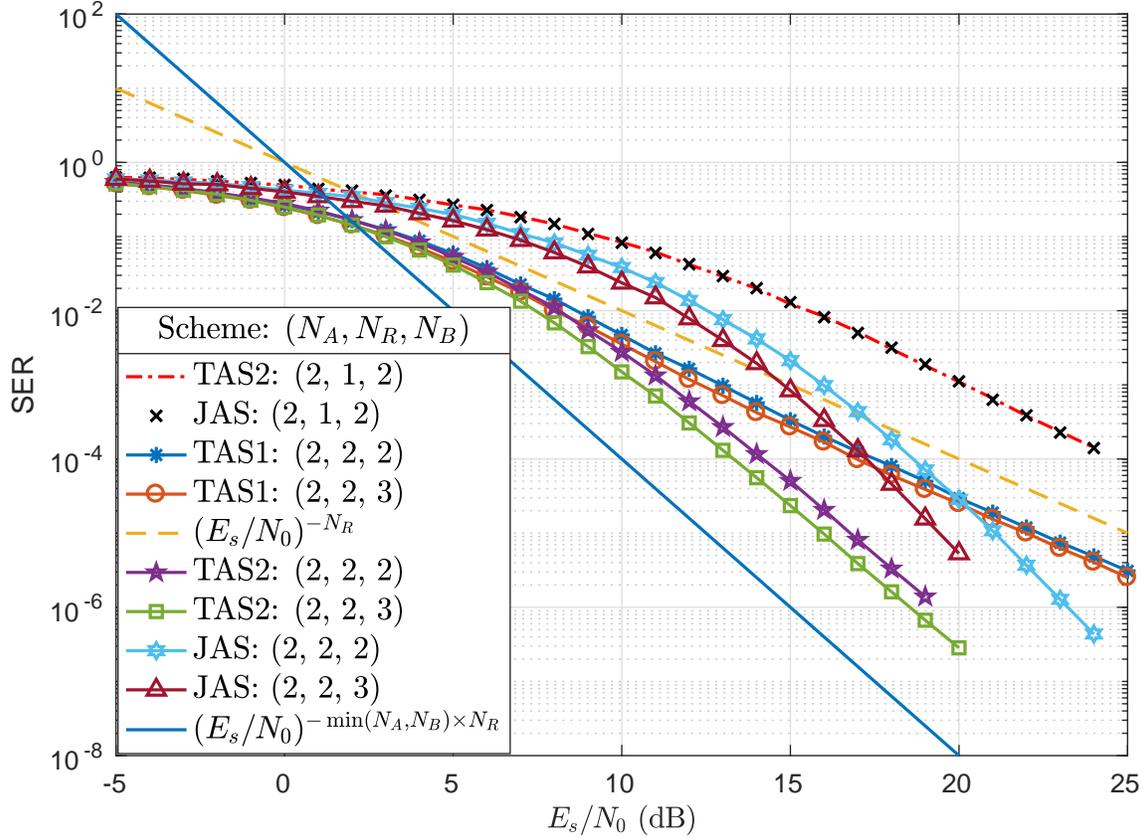}
\caption{Performance comparison of SER for TAS1 and TAS2 in the MA phase for the PNC system.}
\label{MIMO}
\end{figure}
On the other hand, the PNC system with TAS2 achieves a diversity order of $\min(N_{A}, N_{B}) \times N_{R}$ and the SER curve becomes parallel to $(E_{s}/N_{0})^{-\min(N_{A}, N_{B}) \times N_{R}}$ for large $E_{s} / N_{0}$. On the other hand, the PNC system with the JAS scheme also achieves the same diversity order as that of TAS2 with a smaller number of RF chains and results in performance superiority compared to TAS1 for large $E_{s}/N_{0}$ values. 
\section{Conclusions}
The coexistence of a large number of intelligent devices, spectrum scarcity and data-intensive applications in smart cities has attracted the attention of researchers toward new and efficient communication technologies. In this paper, we provide a brief introduction to PNC for bidirectional relaying that has a direct application in WSNs, D2D communications and satellite communications in the context of smart cities. We discussed the distance shortening phenomenon in PNC and suggested simple solutions to mitigate its deleterious effect. In particular, we proposed an ED-based transmit antenna selection scheme as well as a joint user and relay antenna selection scheme, both tailored to the PNC context. With the help of simulations, it was shown that a PNC system with TAS based on the ED metric can achieve the full diversity order of $\min(N_{A}, N_{B}) \times N_{R}$ \emph{without} any need for adaptive network codes and nonstandard modulation schemes, thereby reducing the overall system complexity. Furthermore, simulation results show that the PNC system with JAS can also achieves full diversity order with a smaller number of RF chains compared to ED based TAS. 
\section*{Acknowledgment}
This publication has emanated from research conducted with the financial support from Science Foundation Ireland (SFI) and is co-funded under the European Regional Development Fund under Grant Number 13/RC/2077. \balance

\bibliographystyle{IEEEtran}
\bibliography{PIMRC_SingleColumn}
\end{document}